\documentclass[12pt]{article}
\usepackage{epsfig}
\begin{document}

\begin {center}
{\Large Parity Doubling and highly excited mesons}
\vskip 3mm
{D.V.~Bugg}
\vskip 2mm
{Queen Mary, University of London, London E1\,4NS, UK}
\end {center}

\begin{abstract}
Glozman has proposed that highly excited mesons and baryons fall into
parity doublets, and that the $f_4(2050)$ on the leading Regge
trajectory should have  a nearly degenerate $J^{PC} = 4^{-+}$ partner.
A re-analysis of Crystal Barrel data does not support this idea.
A likely explanation is that centrifugal barriers on the leading
trajectory allow formation of  the $L=J-1$  states,  but are too strong
to allow $L=J$ states.
Two new polarisation experiments have the potential for major
progress in meson spectroscopy.
\end{abstract}

\vskip 2mm

There are two objectives in this Letter.
The first is to report a search for parity doubling
on the leading Regge trajectory for $I=0$, $C=+1$ mesons and relate the
negative result to the centrifugal barrier.
The second objective is to draw attention to the simplicity of
formation experiments compared to production experiments, and point out
that two polarisation measurements have the potential
for major improvements in spectroscopy of light mesons.

Iachello first drew attention to parity doubling in baryon spectra,
i.e. the fact that states with a given $J$ are approximately degenerate
between negative and positive parity \cite {Iachello}.
Glozman pointed out in 2002 that many mesons observed above 1900 MeV
by Crystal Barrel show similar approximate parity doubling
\cite {Glozman}.
He relates this to restoration of chiral symmetry at high excitations.
Glozman and Swanson \cite {Swanson} predicted $J^{PC} = 3^{+-}$
mesons roughly degenerate with $\rho _3(1690)$, likewise $4^{-+}$
degenerate with $f_4(2050)$.
Swanson  extended this prediction to include $3^{++}$ states near 1700
MeV.

Observed states do not presently agree with parity doubling on the
leading (highest) Regge trajectory, though approximate parity doubling
is observed for many states on daughter trajectories.
The well known $\rho_3(1690)$ appears strongly in many sets of data,
but there is no known $3^{+-}$ or $3^{++}$ partner with
isopin $I=0$ or 1 near 1700 MeV.
The high spins of these states should make them conspicuous; all four
appear strongly from 2025 to 2048 MeV, but not at 1700 MeV.
The $a_2(1320)$ is not accompanied by a nearby $2^{-+}$ state;
instead, the $\pi_2(1670)$ appears prominently in many channels as the
lowest $I=1~J^{PC}=2^{-+}$ state.
In Crystal Barrel data, the $f_4(2050)$ appears prominently, but the
lowest observed $4^{-+}$ state is at $2328 \pm 38$ MeV \cite {CBAR}.
Afonin points out similarities
of the observed spectrum to that of the hydrogen atom \cite {Afonin}.

The $\pi$, $\eta$ and $K$ are abnormally light, whereas their
excitations are not, so there is clearly some degree of chiral symmetry
restoration at high mass, though precisely how this works is not yet
agreed.
Before plunging into detail, let us clarify how the partial wave
analysis treats orbital angular momentum $L$, since Glozman argues it
is not a good quantum number.
Glozman argues that the $f_4(2050)$, for example, involves relativistic
quarks obeying the Dirac equation and coupling equally to $^3F_4$ and
$^3H_4$.
However, decays of mesons involve final  states which are not
highly relativistic.
As one example, $\bar pp \to f_4(2050) \to a_2(1320)\pi$, where the
$a_2$ has $\beta = 0.43$.
Decay amplitudes of mesons are written in terms of Lorentz invariant
tensors.
It is necessary to introduce Blatt-Weisskopf centrifugal barriers which
depend on $L$ \cite {Blatt};
explicit formulae for L=1 to 5 are given at the end of Section 2.1 of
Ref. \cite {BSZ}.
The orbital angular momenta are expressed in terms of 3 or 5 powers of
beam momentum, constructed so that tensors for different $L$ are
orthogonal.
A decay amplitude with orbital angular momentum $\ell = 3$
in the final state is constructed likewise in terms of the centre of
mass momentum in the decay and the usual spin 2 tensor for $a_2 \to
\eta \pi$.
The treatment of orbital angular momentum is fully relativistic,
though initial and final states are not highly relativistic.
Decay widths of mesons depend on $L$ and are suppressed for high $L$.

The partial wave analysis includes coupling constants $g$ of $4^+$
states with $L = 3$ and 5 using a fitted ratio
$r_{J=4} = g_{L=5}/g_{L=3}$.
In principle this ratio can have both magnitude and phase.
However, relative phases  are $<20^\circ$ in well determined cases for
all $J$.
The natural interpretatiion  is that a resonance has a unique
phase because of multiple scattering, but the same phase for all decay
channels.
Since phases are consistent with zero, they are set to zero
so as to minimise the number of fitted parameters.

Above a mass of 1900 MeV, there are two complete towers of
$I=0$, $C=+1$, $^{2S+1}L_J$ resonances in eight sets of $\bar
pp$ data, which are fitted with consistent parameters in all channels
\cite {CBAR}.
That analysis was done fitting Crystal Barrel data for
$\bar pp \to \eta \pi^0 \pi^0$, $\pi ^0 \pi ^0$, $\eta \eta$ and
$\eta \eta '$ at nine beam momenta; it included extensive
measurements of both differential cross sections and polarisations
for $\bar pp \to \pi ^+\pi ^-$ from two experiments \cite
{Eisenhandler}  and \cite {Hasan}.
That analysis also fixed the mass and width of $\eta_2 (2267)$ from
Crystal Barrel data on $\bar pp \to \eta ' \pi^0 \pi^0$ \cite
{etaprime}.
Later data on  $\bar pp \to 3\eta$ provided a clear peak  for
$\eta(2320)$ \cite {3eta}.
The present analysis is made to all of these data, fixing masses and
widths of resonances at values from Table 2 of Ref. \cite {CBAR},
except for the $\eta(2320)$.

The polarisation data continue to play a vital role.
They determine imaginary parts of interferences
between triplet partial waves, while differential cross sections
determine real parts of interferences.
This phase sensitivity identifies all $^3H$, $^3F$ and $^3P$ amplitudes
unambiguously, whatever their phases and traces out Argand diagrams.
Polarisation also separates $^3F_2$ and $^3P_2$ mesons cleanly, because
they have orthogonal Clebsch-Gordan coefficients for spin dependent
amplitudes.

This combined analysis has now been rerun, trying to force in a
$4^{-+}$ resonance near 2060 MeV and $5^{++}$ near 2310 MeV.
These are called $\eta_4$ and $h_5$ in the notation of
the Particle Data Group \cite {PDG}.
Formulae are identical to those of Ref. \cite {CBAR}; a more
expansive presentation of formulae is given in my review
paper \cite {Review}.
Breit-Wigner resonances of constant width are fitted to each of 21
$I=0$ resonancs and 6 $I=1$ resonances, (plus tails of two resonances
below the mass range, playing only a small role as backgrounds).
Each resonance requires a complex coupling constant fitted to each
channel of data.
A further detail is that $\pi ^0\pi ^0$, $\eta \eta$ and $\eta
\eta '$ data are fitted to SU(3) formulae \cite {Review}, where $s\bar
s$ components turn out to be small.

There is definite evidence for $\eta _4(2328)$ in $\bar pp \to \eta
\pi ^0 \pi ^0$ data in decays to $[a_0(980)\pi ]_{\ell = 4}$,
$[a_2(1320)\pi ]_{\ell = 2}$ and $[f_2(1275)\eta ]_{\ell =2}$.
Their branching ratios are 1.0:0.28:0.05.
In present data, the $^1G_4$ partial wave interferes only with $\bar pp$
singlet states, since $\bar pp \to \pi \pi$ polarisation involves
only triplet states ($G$ parity $=+1$ for $\pi \pi$).
Two well identified resonances $\eta_2(2267)$ and $\eta(2320)$
interfere with $\eta_4(2328)$ and require that it has resonant phase
variation.
The $\eta_2(2267)$ is one of the most prominent resonances, appearing
as a clear peak in $f_2(1270)\eta '$ \cite {etaprime} and also in
$[f_2(1270)\eta ]_{\ell = 2}$, $[a_0(980)\pi]_{\ell =2}$ and
$[f_0(1500)\eta ]_{\ell = 2}$.
The $\eta (2320)$ appears as a strong peak in $\bar pp \to 3\eta$ in
the $f_0(1500)\eta$ channel \cite {3eta} and also in
$\eta \pi ^0\pi ^0$ data.
The $\eta_4(2328)$ contributes a highly significant improvement
of 558 in log likelihood to the combined analysis.
Log likelihood is defined so that a change of 0.5 is
equal to a change in $\chi^2$ of 1 for the high statistics available.
In assessing the significance level of changes in log likelihood, it is
necessary to use formulae for $\chi^2$ allowing for the number of
degrees of freedom for each resonance.

The full curve of Fig. 1(a) shows the line-shape of $\eta_4(2328)$
normalised to 1 at its peak, which lies at $\sim 2375$ MeV because of
the centrifugal barrier for  production from $\bar pp$.
In searching for the parity partner of $f_4(2050)$,
the $\eta_2(2030)$ and $\eta (2010)$ likewise act as interferometers.
The $\eta_2(2030)$ is visible by eye in Crystal Barrel
data on $\bar pp \to \eta 3\pi ^0$ \cite {2030} and is also required
by $\eta \pi ^0\pi ^0$ data \cite {epp}.
The $\eta(2010)$ of Crystal Barrel is observed in three decays and is
also conspicuous in BES 2 data on $J/\Psi \to \gamma \rho \rho$
\cite {BES2} with a slightly lower mass of 1970 MeV.
\begin{figure} [t]
\begin{center}
\epsfig{file=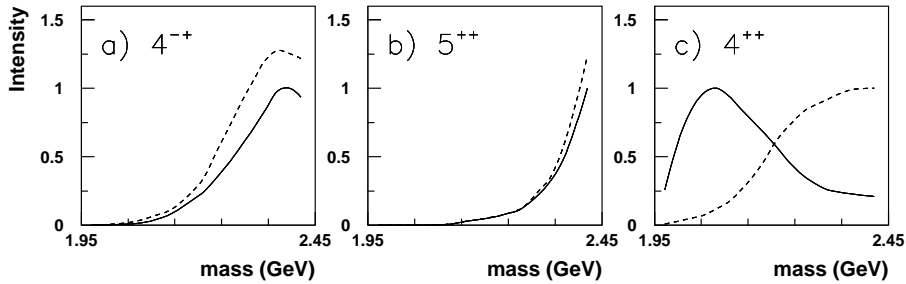,width=14cm} \
\vskip -6mm
\caption{Intensities fitted to (a) $\eta _4(2328)$ alone (full curve)
and the coherent sum of $\eta _4(2328)$ and $\eta _4(2060)$
(dashed curve), (b) likewise for $h_5(2500)$ and $h_5(2310)$,
(c) $f_4(2050)$ (full curve) and $f_4(2300)$ (dashed).}
\end{center}
\end{figure}

If a hypothetical $\eta_4(2060)$ is added to the analysis with
$\Gamma =250$ MeV , decays to $[a_0(980)\pi ]_{\ell = 4}$ and
$[a_2(1320)\pi ]_{\ell=2}$ give small improvements $<40$ in log
likelihood; decays to $[f_2(1275)\eta ]_{\ell=2}$ are negligible.
There is no optimum when the mass and width of $\eta _4(2060)$ are
scanned and it improves log likelihood only by a total of 58.
An improvement $>40$ is normally required for any amplitude to be
regarded as definitive.
The origin of this choice is that there are always some correlations
between $\bar pp$ singlet amplitudes with  $J^{PC} = 0^{-+}$, $2^{-+}$
and $4^{-+}$.
In differential cross-sections, the presence of $4^{-+}$ depends on
terms like $\cos^6 \theta $ and $\cos^8 \theta $, where $\theta$ is
the decay angle of the resonance in its rest frame.
However, there is further confusion from triplet final states with high
spins, which contribute terms in the differential cross section up to
$\cos^8 \theta$.
An extensive simulation shows that a clean identification of an
amplitude requires a change in $ln~L > 40$.
Table 2 of Ref. \cite {CBAR} quotes changes in $ln~L$ for all observed
resonances in $\eta \pi ^0\pi ^0$ data; the $2^+$ states listed there
have much higher log likelihood changes in $\bar pp \to \pi^+\pi^-$.

The dashed curve on Fig. 1(a) shows what
happens: interferences of $\eta _4(2060)$ with $\eta _4(2328)$ enhance
the intensity slightly, but with no significant structure near 2060
MeV.
The likely explanation of the enhancement is that the centrifugal
barriers are slightly wrong.
The Blatt-Weisskopf formula is derived by approximating it with an
equivalent square barrier.
That gives too sharp a rise of the barrier with momentum.
What is desirable is a barrier corresponding to the Coulomb part of the
confining potential with a short-range cut-off or Gaussian smearing.
However, no simple recipe suitable for fitting data exists.
Adding $\eta_4(2060)$ can improve the detailed line-shape of
$\eta_4(2328)$.
Presently, all decays of all resonances are fitted with the same
barrier radius $0.83 \pm 0.03$ fm.
The barrier radius for the $\bar pp$ channel is larger,
$1.11 \pm 0.10$ fm, probably due to the 3 quarks in each nucleon.
If the barrier radii for $J^P=4^-$ are set free, the fit with
$\eta _4(2060)$ alone is midway between full and dashed curves.

A similar test has been made including a hypothetical $^1H_5$ state
$h_5(2310)$.
The logic behind this test is that parity doubling predicts a
$5^{+-}$ partner for $\rho_5(2350)$ which has $J^{PC} = 5^{--}$.
The small mass difference between $u$ and $d$ quarks then predicts a
$5^{++}$ state nearly degenerate with $5^{+-}$.
For $^1H_5$, the situation is less well defined experimentally.
Fig. 1(b) shows as the full curve a fitted $h_5(2500)$ with
$\Gamma = 370$ MeV.
However, this mass is well above the highest experimental data at
2410 MeV, so the `resonance' is simply a parametrisation of the required
$H$ wave.
The dashed curve shows the effect of adding a $h_5(2310)$ with
$\Gamma = 250$ MeV.
Again, constructive interference enhances the fitted signal, but there
is no peak below 2410 MeV where data stop.
Fig. 1(c) shows the line-shapes of $f_4(2050)$ and $f_4(2310)$, which
are observed clearly in $\bar pp \to a_2\pi$ and $f_2\eta$ \cite {epp}.
Note that the $f_4(2050)$ actually has a mass of $2018 \pm 11$ MeV
in the Particle Data Tables. It peaks at 2080 MeV in present
data because of centrifugal barriers.

Table 1 summarises the prominent $^3F$, $^3D_3$ and $^3P_2$ states.
Spin-splitting is mostly tensor and agrees within errors
with that predicted by perturbative QCD; spin-orbit splitting is small.
The $^3D_3$ state at 1982 MeV is particularly prominent in polarisation
data, and is clearly lower than the $F$ states.
The $^3P_2$ state is lower still; this is the $f_2(1910)$ of the
Particle Data group \cite {PDG}.
This pattern is repeated near 2270 MeV.
Splitting between $F$, $D$ and $P$ mesons is consistent with
a stronger centrifugal barrier in $F$ states, which therefore resonate
higher in mass.
A linear extrapolation through $^3P_2$, $^3D_3$ and the centroid of $F$
states predicts a hypothetical $G$ state at $2067 \pm 10$ MeV.
This is rounded down to 2060 MeV to maximise effects of centrifugal
barriers and $^1H_5$ is likewise taken at 2310 MeV.
\begin{table}[htb]
\begin {center}
\begin{tabular}{cccc}
\hline
$^{2S+1}L_J$   & Mass (MeV) & Mass (MeV) \\\hline
$f_4 \equiv ^3F_4$  & $2018 \pm 6$ & $2283 \pm 17$ \\
$f_3 \equiv ^3F_3$  & $2048 \pm 8$ & $2303 \pm 15$ \\
$f_2 \equiv ^3F_2$  & $2001 \pm 10$& $2293 \pm 13$ \\
$h_3 \equiv ^1F_3$  & $2025 \pm 20$& $2275 \pm 25$ \\
$\rho_3 \equiv ^3D_3$  & $1982 \pm 14$& $2260 \pm 20$ \\
$f_2 \equiv ^3P_2$  & $1934 \pm 20$& $2240 \pm 15$ \\\hline

\end{tabular}
\caption{Masses of some $I=0$, $C = +1$ states
from combined Crystal Barrel and PS172 data. }
\end{center}
\end{table}

Could the $\eta _4(2060)$ and $h_5(2310)$ both be invisible because
they are attenuated by centrifugal barriers?
It is instructive to compare the hypothetical $\eta _4(2060)$ and the
$f_4(2050)$, which  is clearly visible with an intensity $10\%$ of
$\eta\pi ^0\pi ^0$ data.
It is necessary to fold the line-shapes of both resonances with
the effects of centrifugal barriers for both $\bar pp$ and decays.
The result is that the $f_4(2050)$ (with mass 2018 MeV) peaks at 2080
MeV and the $\eta(2060)$ would peak at 2150 MeV.
The intensity of a  Breit-Wigner resonance is
\begin {equation} I(s) = \frac {\Gamma_{\bar pp}(s)\Gamma _{decay}(s)}
{|M^2 - s|^2 + |M\Gamma_{total}(s)|^2}
\end {equation}
and $\Gamma _{\bar pp}$ is proportional to the centrifugal barrier.
The ratio $\Gamma _{\bar pp}(4^-)/\Gamma _{\bar pp}(J^P=4^+)$
at their peaks is 0.48.
Conversely, the effect of decay centrifugal barriers is to
enhance the ratio $\Gamma_{decay}(4^-)/\Gamma _{decay}(4^+)$ to 2.4 for
$[a_2\pi ]_{\ell=2}$ at their peaks and to 1.9 for
$[f_2\pi ]_{\ell=2}$.
For decays of $\eta_4(2060)$ to $[a_0(980)\pi ]_{\ell = 4}$, the lower
mass of $a_0(980)$ compared with $a_2(1320)$ almost exactly cancels
the difference between them due to decay barriers.
The  net effect of centrifugal barriers is to reduce the
intensity of $\eta_4(2060)$ by a factor 0.6.
This is significant but not disastrous.
If one makes the assumption that $\eta_4(2060)$ has the same
branching fractions as $\eta_4(2328)$, it should be detectable with an
improvement in log likelihood $>40$ if its central mass is as low as
1980 MeV.
Experimentally, its intensity is no larger than $4\%$ of $f_4(2050)$.

It is of course possible to argue that decay widths may be fortuitously
too weak for these hypothetical states to be detectable.
However, experience for other $J^P$ is that the lowest states
usually appear strongly.
The $f_2(2001)$ (mostly $^3F_2$ in $\bar pp$) and $f_3(2048)$ each
contribute $10\%$ in intensity to $\bar pp \to \eta \pi ^0\pi ^0$;
$h_3(2025) \equiv ^1F_3$ is $5\%$ of $\bar pp \to \omega \eta$.
The $\rho_3(1690)$ and $\pi _2(1670)$ appear prominently in many
channels.
However, it is puzzling that the $\eta _4(2328)$ contributes only
$2.1\%$ of $\eta \pi ^0\pi ^0$ data, while its isospin 1 partner
$\pi_4(2230)$ contributes a huge $59\%$ of all $\bar pp \to \omega \pi
^0$ data.
For both $\pi_4(2230)$ and $\eta_4(2328)$, dominant decays
are with $L=4$ to the lowest available final states $\omega\pi$ and
$a_0(980)\pi$.
This may be attributed to good overlap of wave functions
at the impact parameter of $L=4$ $\bar pp$ interactions.
Then a likely decay mode for $\eta_4(2328)$ (and $\eta _4(2060)$) is
to $[\pi \rho]_{\ell = 4}$, but presently no data are available.

\begin{figure} [t]
\begin{center}
\epsfig{file=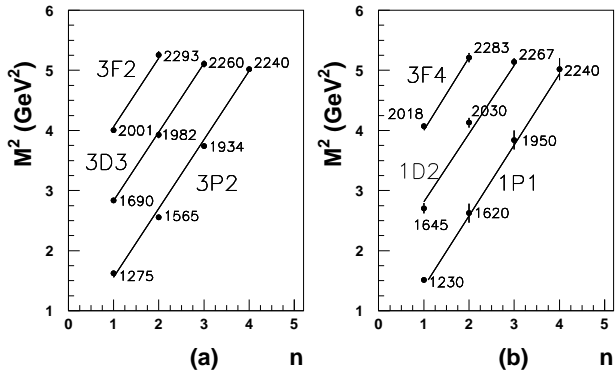,width=10cm} \
\vskip -6mm
\caption{Regge trajectories of some of the $I=0$, $C=+1$ states.}
\end{center}
\end{figure}

Suppose we take at face value the absence of parity partners for
$f_2(1270)$, $\rho_3(1690)$, $f_4(2050)$ and $\rho _5(2330)$.
The Regge trajectories of Fig. 2 show that it costs $\sim
250-300$ MeV for every step in $J$.
The confining potential has a shallow minimum between the centrifugal
barrier at small radius and attraction at long range.
The inference from data is that this shallow minimum is insufficient
to form an $\eta _4(2060)$  resonance with $L=4$.
Important further information from polarisation data is that the
$f_4(2050)$ couples to $\bar pp$ with a ratio of $^3H_4$ and $^3F_4$
waves $r_4 = 0.00 \pm 0.08$; for $f_4(2300)$, $r_4$ rises to
$2.7 \pm 0.5$ \cite {CBAR}.
Still no second $^3H_4$ state appears near 2300 MeV; the first
known $H$ state is $f_6 (^3H_6)$ at $2465 \pm 50$ MeV.
Likewise $\rho _3(1982) (^3D_3)$ has $r_3=0.006 \pm 0.008$ and there is
no evidence for a $^3G_3$ resonance until 2300 MeV.
These results show  that centrifugal barriers suppress states with
$L=J+1$, while allowing  them for $L=J-1$ on the leading trajectory.

If an additional $H$ state orthogonal to $f_4(2050)$ exists near a mass
of 2.1 GeV, it is likely to be narrow.
There is no evidence for narrow peaks in total cross sections, which
have been measured in small steps of mass \cite{Hamilton} \cite {total}.

A different line of argument is that an $\bar nn$ meson may be modelled
as a flux tube with quarks attached to the ends.
Most of the angular momentum is carried by the flux tube, not the
quarks; the angular momentum is then an observable, though relativistic.
The spins of the quarks couple via the Dirac equation at the ends of
the flux tube.
However, to generate a $4^-$ state still requires a flux tube with
$L=4$ compared with $L=3$ for $f_4(2050)$.

In principle a search can also be made for $4^{--}$ states with
$I=0$ and 1 near 2000 MeV.
However, for $C=-1$ states there are no polarisation data, making
separation of $2^{--}$ and $1^{--}$ states difficult and
hindering detection of $4^{--}$ states severely near 2000 MeV.
There is an obvious need for further polarisation data.

Let us now discuss further experiments.
The formation process $\bar pp \to resonance \to decays$ used
by Crystal Barrel is much simpler than production reactions.
The high quality polarisation data on $\bar pp \to
\pi^+ \pi ^-$ demonstrate that it is a practical proposition to find a
complete spectrum of $I=0~C=+1$  states from 1910 to 2400 MeV.
The $I=1~C=-1$ spectrum is almost complete, but $^3S_1$ and $^3D_1$
states are poorly identified.
The phase sensitivity available from polarisation data is absent in
production experiments; as a result, only strong resonances appearing
as peaks or states interfering with well established resonances can
be identified.
Polarisation data are crucial, as in baryon spectroscopy.

Two key experiments are needed, each with a $4\pi$ detector, good
$\gamma$ detection and transverse polarisation.
At FLAIR, the $\bar p$ ring under construction at GSI with
momenta up to 2 GeV/c, measurement of polarisation from 360 to 1940
MeV/c for neutral final states would be straightforward in detectors
like Belle, Babar and Cleo C, which are all now idle; the detector does
not need a magnetic field.
The low momenta are crucial in probing the lower
sides of resonances clustered near 2000 MeV.

I have made a Monte Carlo simulation based on existing differential
cross sections and the Crystal Barrel detector as an example \cite
{Experiments}.
It shows that polarisation data for $\bar pp \to \omega \pi ^0$,
$\omega \eta$ and $\omega \eta \pi ^0$ (detecting $\omega$ in
$\pi^0\gamma$ as in Crystal Barrel) could complete the $I=1~C = -1$
and $I=0~C -1$ states; they would also allow checks on parity doubling
of $a_4(2040)$.
Data for $\eta \pi ^0$ and $\eta \eta \pi ^0$ would probably complete
the $I=1$ $C=+1$ states, which already resemble $I=0$ $C =+1$, but with
one discrete ambiguity in the $\eta \pi$ solution.
They would allow tests for $4^{--}$ states near $f_4(2050)$.
Polarisation for $\bar pp \to \eta \pi^0 \pi ^0$ would check the
present $I=0$ $C=+1$ spectrum.
They would provide interferences between singlet $\bar pp$ states and
known triplet states, hence improving the work presented here.

The technology of polarised targets is well developed and
costs are modest.
Background from nitrogen nuclei in an ammonia target are
estimated by the Monte Carlo simulation to be $\le 10\%$ after kinematic
fitting, and are of similar magnitude to cross-talk between final
states.
That has been demonstrated in an experiment at LAMPF using a polarised
beam and polarised target to study $pp \to pn \pi ^+$ \cite {LAMPF}.
The Fermi momentum in the nucleus is $\sim 120 $ MeV/c along each
of $x$, $y$ and $z$ axes, compared with errors $\le 20$ MeV/c for
reconstructed photons.
Seven of the nine momenta used by Crystal Barrel with $\bar p$ in
flight were taken in 4 calendar months, so running time is quite
reasonable for a high return of physics.

A related programme at VEPP2 \cite {VEPP2} and VEPP4 \cite
{VEPP4} (Novosibirsk) using CMD2 and SND detectors and transversely
polarised electrons could separate $^3S_1$ and $^3D_1$ amplitudes.
Transverse polarisation is a linear combination of $s_z =\pm 1$.
For  $^3D_1$ components, they produce highly distinctive azimuthal
asymmetries of the form $\cos \phi$ and $\cos 2\phi$, where $\phi$ is
the azimuthal angle between the final state and the plane defined by
the initial polarisation and the beam.
Clean identification of  $J^P=1^{--}$ states and the ratios of $^3D_1$
and $^3S_1$ amplitudes would provide quantitative information relevant
to chiral symmetry restoration.
If $1^{--}$ states follow the pattern of other $J^{PC}$, a
$^3S_1$ state recurrence of the $\rho$ is expected near 1300 MeV and
two states $^3S_1$ and $^3D_1$ near 1700 MeV.
Above this the experimental situation is confused.
Information on $I=0$ and $s\bar s$ states is scanty.
Such data above 1900 MeV would improve greatly the
analysis of Crystal Barrel data with $C=-1$.

Those readers interested in the existing data should consult my
review \cite {Review}, where a full set of references and listings
of all observed decay modes are given in Tables 2, 5, 7 and 8.
Listings of the Particle Data Group do not include observed decay
channels nor complete lists of the reactions studied.
Only final combined analyses of all data are referenced.
For $I=1$, $C=+1$, tabulations are out of date (despite promptings)
and the final combined analysis \cite {I1} is not referenced nor masses
and widths determined there.
The casual reader could be misled into believing that the listings of
Crystal Barrel states are single observations of each resonance.
In fact, the great majority have been observed in two or more decays
and several in four or five decays and up to 8 sets of data.

Clarification of the spectroscopy of light mesons and baryons is
crucial to a full understanding of QCD and confinement, which is one of
the basic phase transitions of physics.

\end {document}